\newcommand{\be}{\begin{equation}}
\newcommand{\ee}{\end{equation}}
\documentstyle[12pt]{article}
\begin{document}
\thispagestyle{empty}
\begin{center}
\large An Heterotic SUSY Version\\
of one Non--Stable Integrability\\
Theorem Due to K.H. Mayer\\
\end{center}
\vskip1cm
\centerline{\bf Juan Fernando Ospina G.}
\vskip.5cm
\begin{center}
Departamento de F\'\i sica\\
Universidad de Antioquia\\
A.A. 1226, Medellin, Colombia
\end{center}
\vskip1cm
\begin{abstract}
By using Heterotic SUSY path integral, one version and ge\-ne\-ra\-li\-
za\-tion of an integrability theorem due to Mayer is obtained.
\end{abstract}
\vskip3truecm

\vfill\eject
\section{Introduction}
The typical relation which links heterotic SUSY with differential topology
is~\cite{k:ch}
\[
\hbox{``Heterotic SUSY Path Integral}=\hbox{integer"}
\]
\be
\int_{\rm H.SUSY.\ B.C.}{\cal D}{\cal M}_F(\hbox{Heterotic 
SUSY})e^{-S(\rm Heterotic \ SUSY)}=\hbox{integer}
\label{1.1}
\ee
${\cal D}{\cal M}_F$ (Heterotic SUSY) is an heterotic SUSY-Feynmann
integration measure which splits in bosonic and heterotic fermionic parts:
\be
{\cal D}{\cal M}_F(\hbox{Heterotic SUSY})={\cal D}{\cal 
M}_F)(\hbox{Bosonic})\ {\cal D}{\cal M}_F(\hbox{Heterotic Fermionic})
\label{1.2}
\ee
$S$(Hetrotic SUSY) is a heterotic SUSY action. (H. SUSY. B. C) is a 
heterotic set of SUSY boundary conditions:
\be
(\hbox{H. SUSY. B. C})=(\hbox{B. B. C})(\hbox{H. F. B. C})
\label{1.3}
\ee
(B. B. C) is a set of bosonic boundary conditions. (H. F. B. C) is a 
heterotic set of fermionic boundary conditions. (\ref{1.1}) can be written as
(using (\ref{1.2}) and (\ref{1.3}):
\begin{eqnarray}
{\int_{(\rm B. B. C)(\rm H. F. B. C)}{\cal D}{\cal
M}_F(\hbox{Bosonic}){\cal D}{\cal M}_F(\hbox{Heterotic 
Fermionic})e^{-S({\rm heterotic \ SUSY})}} && \nonumber\\
= \hbox{integer} &&
\label{1.4}
\end{eqnarray}
For to evaluate the heterotic SUSY path integral of (\ref{1.4}) one must give a
heterotic SUSY lagrangian.

\section{The Lagrangian}

\begin{eqnarray}
{\cal L} & = & \alpha g_1(\dot \Phi, \dot \Phi)+ \beta g_1(\Psi, \nabla_{\dot
\Phi}\Psi)+\sum_{i=1}^4 \beta_i g_i(\Psi_i, \nabla_{\dot 
\Phi}^i\Psi_i) \nonumber \\
& + & \sum_{i=1}^3 \gamma_i F_i(\Psi_{i+1}, \Psi_{i+1}, 
\Psi_1, \Psi_1)
\label{2.1}
\end{eqnarray}
$\{g_i\}$ is a set of metrics. $\{F_i\}$ is a set of curvatures of 
Yang-Mills field. $\Phi$ is a set of bosonic coordinates. $\{\Psi, 
\{\Psi_i\}\}$ is a set of fermionic fields. $\{\alpha, \beta, \{\beta_i\}, 
\{\gamma_i\}\}$ is a set of constants. $\dot \Phi$ is the derivate of 
$\Phi$ with respect to the time. $\{\nabla_{\dot \Phi}$, $\{\nabla_{\dot 
\Phi}^i\}\}$ is a set of covariant derivates with respect $\dot \Phi$.

Taking $\Phi$ like a perturbation $U$ with respect to a background 
$\Phi_0$ ($\Phi =\Phi_0 + U$). Taking $\Psi_1$ like a perturbation with 
respect to a background $\Psi_0$: \cite{kob:fu} $\Psi_1=A_U\Psi_0$.
$A_U=L_U-\nabla_U$; $L_U$ is the Lie derivate with respect to $U$. Taking
$\Psi_3$, $\Psi_4$ paraleles (constants) and using $\nabla_Y(A_X)=R(X, 
Y)$; (\ref{2.1}) transforms to:
\begin{eqnarray}
{\cal L} & = & \alpha g_1(\dot U, \dot U)+\beta g_1(\Psi, \nabla_{\dot 
U}\Psi)+\beta_1 g_1(\Psi_0, R(U {\dot U})\Psi_0) \nonumber \\
 & + & \beta_2 g_2(\Psi_2, \nabla_{\dot U}^2\Psi_2)
+\sum_{i=1}^3 \gamma_i F_i(\Psi_{i+1}, \Psi_{i+1}, \Psi_0, \Psi_0)
\label{2.2}
\end{eqnarray}
In tensorial notation (\ref{2.2}) can be written as:
\begin{eqnarray}
{\cal L} & = & \alpha g_{ab}{\dot U^a}{\dot U^b}+\beta 
g_{ab}\Psi^a\nabla_{\dot U}\Psi^b +\beta_1 R_{abcd}\Psi_0^c\Psi_0^d 
U^a{\dot U^b} \nonumber \\ 
& + & \beta_2 g_{AB}\Psi_2^A\nabla_{\dot U}\Psi_2^B
+\gamma_1 F_{1abAB}\Psi_0^a\Psi_0^b\Psi_2^A\Psi_2^B \nonumber \\
& + & \gamma_2 
F_{2ab\alpha \beta}\Psi_0^a\Psi_0^b\Psi_3^\alpha \Psi_3^\beta +\gamma_3 
F_{3ab\mu\nu}\Psi_0^a\Psi_0^b\Psi_4^\mu\Psi_4^\nu
\label{2.3}
\end{eqnarray}
Where small latin indexes correspond to bosonic target space; capital 
indexes correspond to the group of the Yang-Mills field $F_1$; greek indexes
correspond to the group of the Yang-Mills field $F_2$ (the first ones) and the 
Yang-Mills field $F_3$ (the last ones).

\section{The Integration Measure}
Tha bosonic measure is 
\[
{\cal D}{\cal M}_F(\hbox{Bosonic})={\cal D}\Phi_0{\cal D}U
\]
The heterotic fermionic measure is
\[
{\cal D}{\cal M}_F(\hbox{Heterotic Fermionic})={\cal D}\Psi_0{\cal 
D}\Psi_2{\cal D}\Psi_3{\cal D}\Psi_4{\cal D}\Psi
\]
(\ref{1.2}) can be written as:
\be
{\cal D}{\cal M}_F(\hbox{Heterotic SUSY})={\cal D}\Phi_0{\cal D}U {\cal
D}\Psi_0{\cal D}\Psi_2{\cal D}\Psi_3{\cal D}\Psi_4{\cal D}\Psi
\label{3.1}
\ee 

\section{Boundary Conditions}
(\ref{1.3}) can be written as:
\be
(\hbox{H. SUSY. B. C})=(\hbox{B. B. C})(\hbox{H. F. B. C}) =(\hbox{B. C})_U
(\hbox{B. C})_\Psi (\hbox{B. C})_{\Psi_2}
\label{4.1}
\ee

\section{The Calculation of the Path Integral}
Using (\ref{2.3}), (\ref{3.1}), (\ref{4.1}) and assuming that $\nabla_{\dot U}\simeq
\delta_t$ and $g\simeq \delta$; (\ref{1.4}) transforms to: ( \( S=\int {\cal
L}dt \) )
\begin{eqnarray}
& & \int {\cal D}\Phi_0{\cal D}\Phi_0 \left[ \int_{\rm (B. C)_{\Psi}} {\cal
D}\Psi e^{-\beta \int \delta_{ab}\Psi^a{\dot \Psi^b}dt} \right] 
\cdot \left[ 
\int_{\rm (B.C)_U}{\cal D}Ue^{-\int(\alpha \delta_{ab}{\dot U^a}{\dot 
U^b}+\beta_1 R_{abcd}\Psi_0^c\Psi_0^d U^a{\dot U^b})dt} \right] \cdot
\nonumber \\
& & \left[ \int_{\rm (B. C)_{\Psi_2}}{\cal D}\Psi_2 e^{-\int (\beta_2 
\delta_{AB} \Psi_2^A{\dot \Psi_2^B}+\gamma_1
F_{1abAB}\Psi_0^a\Psi_0^b\Psi_2^A\Psi_2^B)dt} \right ] \cdot \left[
\int {\cal D}\Psi_3 e^{-\int \gamma_2 F_{2ab\alpha \beta}\Psi_0^a\Psi_0^b 
\Psi_3^\alpha \Psi_3^\beta dt} \right ] \cdot \nonumber \\
& & \left[\int {\cal D}\Psi_4 
e^{-\int \gamma_3 F_{3ab\mu\nu}\Psi_0^a\Psi_0^b\Psi_4^\mu\Psi_4^\nu 
dt} \right ] \nonumber \\
& & =\hbox{integer}
\label{5.1}
\end{eqnarray}
The regularization of (\ref{5.1}) is:
\be
\int_{\rm M}\left [\hat A(R)\prod_{i=1}^S 
2\cosh\left(\frac{y_i}{2}\right)ch(F)e^{\frac{d}{2}}\right ]_{\rm Top-form}=
\hbox{integer}
\label{5.2}
\ee
$\{y_i\}$ is the set of eigenvalues of $F_1$ (matrix $2S \times 2S$).
$\hat A(R)$ is de Dirac genus. $ch(F)$ is the Chern character for the 
Yang-Mills $F_2$. $e^{d/2}$ is the cohomological term for the Yang-Mills 
$F_3$. $M$ is the bosonic space.

\section{Mayer's Theorem}
(\ref{5.2}) is a integrality theorem for the manifold $M$. (\ref{5.2}) is the
integrality theorem due to Mayer trated by Hirzebruch. In fact:
\cite{hi:to}

Considering that Yang-Mills $F_1$ corresponds to a SO(2S)-bundle over $M$
with formal factorisation \( p(F_1)=\prod_{i=1}^S (1+y_i^2) \). 
Considering that Yang-Mills $F_2$ corresponds to a $GL(q, {\bf
C})$-bundle over $X$. Considering that Yang-Mills $F_3$ induces an element 
$d\in H^2(M, {\bf Z})$ whose reduction $mod\ 2$ is $w_2(X)+w_2(F_1)$. 
Then (\ref{5.2}) is the expresion of cohomological quantization for $M$.
(\ref{5.2})
is the Mayer's theorem.

\section{Conclusions}
(\ref{5.2}) was obtained using ${\rm (B. C)_{\psi_2}}$; if one use another
boundary conditions for $\Psi_2$, to say, ${\rm (B. C)}'_{\Psi_2}$; one obtain:
\be
\int_{\rm M}\left [ A(R) \prod_{i=1}^S \sinh \left( \frac{y_i}{2}\right 
)ch(F) e^{\frac{d}{2}} \right]_{\rm Top-form}=\hbox{integer}
\label{6.1}
\ee
(\ref{5.2}) and (\ref{6.1}) have the following generalizations: \cite{k:ch}
\be
\int_{\rm M}\left[ A(R)\prod_{i=1}^S 2\cosh\left(\frac{y_i}{2}\right)ch(F)
e^{\frac{d}{2}}(\hbox{string structure})\right ]_{\rm 
Top-form}=\hbox{character}
\label{6.2}
\ee
\be
\int_{\rm M}\left[ A(R)\prod_{i=1}^S 
2\sinh\left(\frac{y_i}{2}\right)ch(F) e^{\frac{d}{2}}(\hbox{string
structure})\right]_{\rm Top-form}=\hbox{characte}
\label{6.3}
\ee

\end{document}